\documentclass[a4paper]{article}

\usepackage{INTERSPEECH2020}
\usepackage[utf8]{inputenc}
\usepackage{enumitem}
\usepackage{graphicx}
\usepackage{multirow}
\usepackage{verbatimbox}

\usepackage[hyphens]{url}
\usepackage[hidelinks]{hyperref}
\hypersetup{breaklinks=true}
\usepackage{cite}

\title{Streaming keyword spotting on mobile devices}
\name{Oleg Rybakov$^1$, Natasha Kononenko$^1$, Niranjan Subrahmanya$^2$, Mirkó Visontai$^2$, Stella Laurenzo$^1$}
\address{$^1$Google Research  $^2$Google Speech}
\email{\{rybakov, natashaknk, sniranjan, mirkov, laurenzo\}@google.com}

\begin{document}

\let\<\textless
\let\>\textgreater

\maketitle
\begin{abstract}
In this work we explore the latency and accuracy of keyword spotting (KWS) models in streaming and non-streaming modes on mobile phones. NN model conversion from non-streaming mode (model receives the whole input sequence and then returns the classification result) to streaming mode (model receives portion of the input sequence and classifies it incrementally) may require manual model rewriting. We address this by designing a Tensorflow/Keras based library which allows automatic conversion of non-streaming models to streaming ones with minimum effort. With this library we benchmark multiple KWS models in both streaming and non-streaming modes on mobile phones and demonstrate different tradeoffs between latency and accuracy. We also explore novel KWS models with multi-head attention which reduce the classification error over the state-of-art by 10\% on Google speech commands data sets V2. The streaming library with all experiments is open-sourced.
\footnote{Preprint. Submitted to INTERSPEECH.} % TODO remove it
\end{abstract}

\section{Introduction}

Research and development of neural networks has many steps: data collection, model design and training, model latency or memory footprint optimization, model conversion to inference mode and execution of the model on different hardware, including mobile devices. In this work we are focused on the last three steps: model optimization, model conversion to inference mode and running it on mobile devices. A common method of model optimization is quantization \cite{QUANT1, QUANT2} allowing to reduce both model size and latency. It can be applied to many problems, including computer vision \cite{QUANT1} and speech recognition \cite{STREAM_ASR}. Another optimization approach is the transformation of the NN model (which was used during training) into another NN streaming model which can be more efficient for inference/prediction mode \cite{STREAM_CONV}. In several applications, such as image classification, model representation in the training and inference modes are the same, while in others, such as sequence classification problems (for example, KWS), it can be different.

\subsection{Model streaming example}

\begin{figure}[t]
  \centering
  \includegraphics[width=\linewidth]{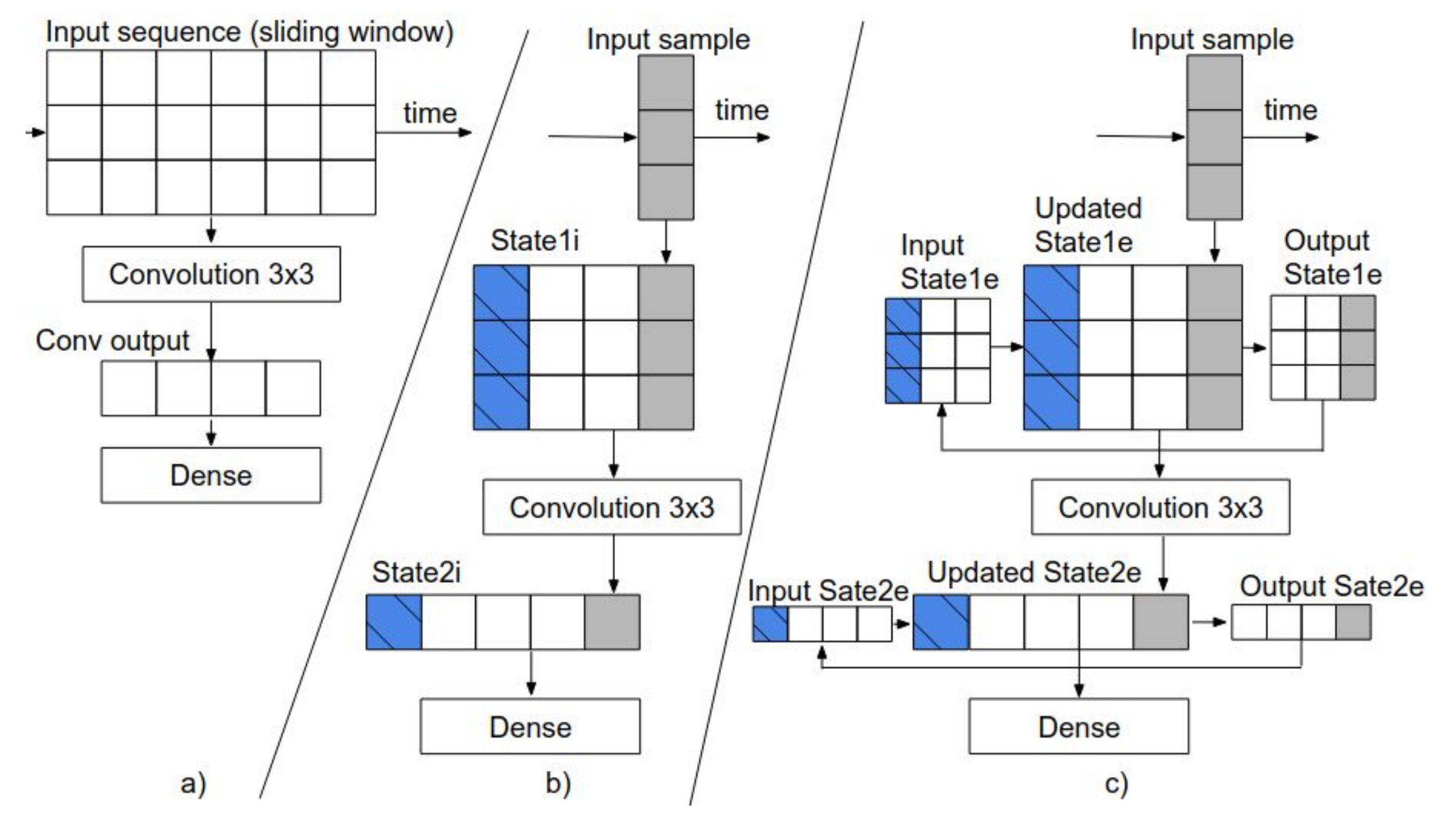}
  \caption{Convolutional model: a) non-streaming b) streaming with internal state c) streaming with external state}
  \label{fig:streaming_convolution}
  \vspace{-5mm} % TODO remove it
\end{figure}

Let’s consider an example of convolutional NN (shown on Fig~\ref{fig:streaming_convolution}a) applied on KWS. A standard approach for model training is to use a non-streaming model representation, shown on Fig~\ref{fig:streaming_convolution}a. It receives the whole input sequence and then returns the classification result (on Fig~\ref{fig:streaming_convolution} the whole input sequence has length 6 with single sample feature size 3).

In a KWS application we do not know when the keyword starts or ends, so we need to process every audio packet and return the classification results in real time every 20ms (for example) - it is called streaming inference. It will not be efficient to use a non-streaming model representation (Fig~\ref{fig:streaming_convolution}a) in streaming inference mode because the same convolution will be recomputed on the input window multiple times. A standard approach to optimize latency in this case is to transform the non-streaming model (Fig~\ref{fig:streaming_convolution} a) to a streaming one \cite{STREAM_CONV}. The resulting model will receive input samples with dimensions 3x1 incrementally process it (keeping the previously computed convolutions in a buffer to avoid unnecessary computations) and return the classification results in a streaming fashion, as shown on Fig~\ref{fig:streaming_convolution}b, c.

There are two options of implementing streaming inference: with internal state (Fig~\ref{fig:streaming_convolution}b) and with external state (Fig~\ref{fig:streaming_convolution}c). A model with internal state receives a new input sample with dimensions 3x1 (grey box 3x1 on Fig~\ref{fig:streaming_convolution}b), appends it to the ring buffer State1i, and at the same time removes the oldest sample (marked by blue with dashed lines on State1i) from State1i. This way, State1i always has a shape of 3x3. In the next step State1i is used by convolution 3x3 (on Fig~\ref{fig:streaming_convolution}b). Then conv output (1x1 grey box) is concatenated with buffer State2i. At the same time we remove the oldest sample (marked by blue with dashed lines on State2i) from State2i. In the end, State2i is fed into a dense layer. As a result, the convolution will be computed on the new input data only and all previous computations will be buffered, as shown on Fig 1 b. In this case states are internal variables of the model which have to be updated with every prediction. The same approach is shown on Fig~\ref{fig:streaming_convolution}c, with one difference: states State1e and State2e are inputs and outputs of the model. In this case model does not keep any internal states and developers will have to feed them into neural network as additional inputs and then in addition to classification results receive the updated states and feed them back on the next prediction cycle.

In the above example, model representation in training and inference models can be different, so developers have to reimplement the NN model in streaming mode (for model latency optimization). In this work we would like to automate it, so that developers could write a NN model once, train it and then automatically convert it to streaming mode. For this purpose we designed a Keras streaming wrapper layer, described in section 2.

\subsection{About speech frontend and modeling pipeline}

KWS NN models use a speech feature extractor based on MFCC \cite{MFCC}. Our implementation of speech feature extractor supports both FFT and DFT. If DFT is selected, then it will increase model size, because of DFT weights. We implemented all the components of the speech feature extractor in Keras layers, so that they will be part of the Keras NN model and will be automatically converted to TFLite \cite{TFLITE}. Since the speech feature extractor is a part of the model it will be easier to deploy it on a mobile device. The overall modeling pipeline (with given data)  will have several steps:

\begin{itemize}[noitemsep]
\item Design a model using Keras layers and Stream wrapper.
\item Train the model.
\item Automatically convert the trained non-streaming Keras model to Keras streaming one.
\item Convert streaming Keras model to a TFLite module \cite{TFLITE}. Quantize it if needed to optimize model size and performance.
\item Benchmark TFLite module: measure accuracy and latency on CPU of mobile phone Pixel4 \cite{PIXEL4}.
\end{itemize}

\noindent The main contributions of this paper are:

\begin{itemize}[noitemsep]
\item We implemented several popular KWS models in Keras and designed a streaming Keras layer wrapper for automatic model conversion to streaming inference with internal/external states.
\item We improved the classification error of the state of the art KWS model by 10\% on datasets V2 \cite{DATA2}.
\item We trained KWS NN models on Google speech commands dataset \cite{DATA} and compared their accuracy with streaming and non-streaming latency on a Pixel 4 phone.
\item The code, pretrained models and experimental results are open-sourced and available at \cite{OPEN}.
\end{itemize}

\section{Model streaming}

We designed a streaming Keras layer wrapper which allows automatic conversion of Keras models to streaming inference (with internal or external states, Fig~\ref{fig:streaming_convolution}b, c). With this approach developer does not need to manually rewrite the model for streaming mode. This allows us to reduce the time to model deployment. Below is an example of a functional Keras model:

\begin{verbnobox}[\fontsize{7pt}{7pt}\selectfont]
output = tf.keras.layers.Conv2D(...)(input)
output = tf.keras.layers.Flatten(...)(output)
output = tf.keras.layers.Dense(...)(output)
\end{verbnobox}

\noindent By wrapping layers (which have to be streamed and require a buffer as shown in Fig~\ref{fig:streaming_convolution}b, c) with the Stream wrapper we get the model (in this example Dense layer does not keep any states in time, so it is streamable by default):

\begin{verbnobox}[\fontsize{7pt}{7pt}\selectfont]
output = Stream(cell=tf.keras.layers.Conv2D(...))(input)
output = Stream(cell=tf.keras.layers.Flatten(...))(output)
output = tf.keras.layers.Dense(...)(output)
\end{verbnobox}

\noindent Now we can train this model (with no impact on training time) and convert it to streaming inference mode automatically.
Stream wrapper creates states and manages in inference mode. In addition, the Stream wrapper needs to know the effective time filter size which is different for different layers, that is why inside of the Stream wrapper we have different logic for extracting the effective time filter size for a particular layer.

We designed the Stream wrapper with several requirements. First, it shouldn't impact the original model training or default non-streaming inference and will be used only for streaming inference. In training mode it uses the cell as it is, so the training time will be the same as without the Stream wrapper. Next, we would like to support cells with internal and external states as it is shown in Fig~\ref{fig:streaming_convolution}b (with internal state) and Fig~\ref{fig:streaming_convolution}c (with external state). So that we can use different inference engines and compare them with each other.

RNN layers also require states during streaming inference, so we build streaming-aware RNN layers with streaming function. It behaves as standard RNN during training, but after model conversion to streaming inference it will only call the RNN cell (with internal or external state defined by the user).

Automatic conversion to streaming inference mode has several steps: 1) set input layer feature size equal one frame; 2) traverse Keras NN representation and insert ring buffer for layers which have to be streamed or call streaming function in streaming-aware layers such as RNN. This approach is not specific to KWS models and can be used in other applications.

Current version of Stream wrapper does not support striding nor pooling more than 1 in the time dimension, but it can be implemented in the future.

\section{Model architectures}

The standard end to end model architecture applied for KWS \cite{DNN1, CNN2} consists of a speech feature extractor(optional) and a neural network based classifier. An input audio signal of length 1sec is framed into overlapping frames of length 40ms with an overlap of 20ms as in \cite{HELLO}. Each frame is fed into a speech feature extractor which is based on MFCC \cite{MFCC}. The extracted features are classified by a neural network which produces the probabilities of the output classes. We use cross entropy loss function with Adam optimizer for model training. In this section we overview the neural network architectures evaluated in this work, We implemented popular models from \cite{HELLO},\cite{ATT2},\cite{SVDF1}, in our library for benchmarking streaming and non streaming inference on mobile phone.

\subsection{Deep Neural Network (DNN)}
The DNN model applies fully-connected layers with rectified linear unit (ReLU) activation function on every input speech feature. Then the outputs are stacked over 49 frames and processed by a pooling layer followed by another sequence of fully-connected layers with ReLU. We observed that this architecture with a pooling layer gives higher accuracy than the standard DNN model published in \cite{HELLO}, as shown in Table~\ref{tab:accuracy}. We use a similar number of model parameters with DNN model in \cite{HELLO}.

\subsection{Convolutional Neural Network (CNN)}
CNN\cite{CNN1} is a popular model which is used in many applications including KWS \cite{CNN1, CNN2, TEMPCONV}. As in \cite{HELLO} it is composed of sequence of 2D colvolutions with ReLU non linearities followed by fully connected layers in the end. Below we benchmarked several variations of CNN: one with striding equal 2 (CNN+strd on Table~\ref{tab:accuracy} and Table~\ref{tab:accuracy_latency_size}) and another with no striding (CNN on Table~\ref{tab:accuracy_latency_size}, it can be automatically converted to streaming mode).

\subsection{Recurrent Neural Networks: LSTM, GRU}
RNNs\cite{LSTM1} are successfully applied in KWS problems \cite{CRNN, LSTM5}. It receives speech features as a sequence and processes it sequentially with updating the internal states of the model. This approach is well suited for streaming inference mode. We explore two versions of RNN: LSTM \cite{LSTMP} (in Table~\ref{tab:accuracy} called LSTM) and a GRU-based model \cite{GRU1} and compare them with baseline \cite{HELLO} on Table~\ref{tab:accuracy}.

\subsection{Convolutional Recurrent Neural Network (CRNN)}
CRNN\cite{CRNN} combines properties of both CNN, which captures short term dependencies, and RNN, which uses longer range context. It applies a set of 2D convolutions on speech features followed by a GRU layer with fully connected layers. We compare CRNN model in our library with baseline \cite{HELLO} on Table~\ref{tab:accuracy}.

\subsection{Depthwise Separable Convolutional Neural Network (DSCNN)}
DSCNN\cite{DS_CNN} models are well applied on KWS \cite{HELLO}. This model processes speech features by using a sequence of 2D convolutional and 2D depthwise layers followed by batch normalization with average pooling (as in \cite{HELLO}) and finished by applying fully connected layers. Below we benchmarked several variations of DSCNN: one with striding equal 2 (it is similar to DSCNN in \cite{HELLO} and shown as DSCNN+strd on Table~\ref{tab:accuracy} and Table~\ref{tab:accuracy_latency_size}) and another with no striding (DSCNN on Table~\ref{tab:accuracy_latency_size}, it can be automatically converted to streaming mode).

\subsection{Multihead attention RNN (MHAtt-RNN)}
The development of the Attention mechanism \cite{ATT0, ATT1} improved the accuracy on multiple tasks including KWS\cite{ATT2}. In \cite{ATT2} the authors build a model Att-RNN which takes a mel-scale spectrogram and convolves it with a set of 2D convolutions. Then two bidirectional LSTM \cite{LSTM1} layers are used to capture two-way long term dependencies in the audio data. The feature in the center of the bidirectional LSTM’s output sequence is projected using a dense layer and is used as a query vector for the attention mechanism. Finally, the weighted (by attention score) average of the bidirectional LSTM output is processed by a set of fully connected layers for classification \cite{ATT2}. We extended this approach with multi-head attention (4 heads) and replaced LSTM with GRU (and call this version MHAtt-RNN). It allows us to reduce classification error by 10\% in comparison to the state of the art (shown in Table~\ref{tab:accuracy}). Both Att-RNN and MHAtt-RNN models are using bidirectional RNN, so they cannot be converted to streaming mode and have to receive the whole sequence before producing classification results.

\subsection{Singular value decomposition filter (SVDF)}
We implemented a simplified version of \cite{SVDF1}, so that it does not require aligned annotation of audio data for training. This model is composed of several SVDF and bottleneck layers with one softmax layer in the end. The SVDF block is a sequence of one dimensional convolution and one dimensional depthwise convolution layers\cite{SVDF1}.

\subsection{Temporal Convolution ResNet (TC-ResNet)}
The TC-ResNet model\cite{TEMPCONV} is composed of sequence of residual blocks which use one dimensional convolution. In this paper we use neural network topology called TC-ResNet14\cite{TEMPCONV}. To improve accuracy of this model we increase number of parameters from 305K to 365K and use SpecAugment \cite{SPEC}. Baseline TC-ResNet accuracy with its improvement on data sets V1 and V2 are shown on Table~\ref{tab:accuracy}.

\begin{table}[t]
  \caption{Baseline models accuracy on data V1* and V2* with paper references and our models accuracy on data V1 and V2}
  \label{tab:accuracy}
  \centering
  \begin{tabular}{p{1.8cm} p{1.2cm} p{0.9cm} p{1.2cm} p{0.9cm}}
    \toprule
    \textbf{Models} & \textbf{V1*[\%]} & \textbf{V1[\%]} & \textbf{V2*[\%]} & \textbf{V2[\%]}  \\
    \midrule
      DNN        & 86.7\cite{HELLO}    & 91.2 &      &  90.6  \\
      CNN+strd   & 92.7\cite{HELLO}    & 95.4 &      &  95.6  \\
      SVDF       &                     & 96.3 &      &  96.9  \\
      DSCNN+strd & 95.4\cite{HELLO}    & 97.0 &      &  97.1  \\
      GRU        & 94.7\cite{HELLO}    & 96.6 &      &  97.2  \\
      LSTM       & 94.8\cite{HELLO}    & 96.9 &      &  97.5  \\
      CRNN       & 95.0\cite{HELLO}    & 97.0 &      &  97.5  \\
      Att-RNN    & 95.6\cite{ATT2}     &      & 96.9\cite{ATT2} &        \\
      TC-ResNet  & 96.6\cite{TEMPCONV} & 97.1 &      &  97.4      \\
      Embed+head &                     &      & 97.7\cite{LIMSDATA} &        \\
      MHAtt-RNN  &                     & 97.2 &      &  98.0  \\
    \bottomrule
  \end{tabular}
  \vspace{-5mm} % TODO remove it
\end{table}

\section{Experimental results}

\subsection{Datasets and accuracy metrics}

On Table~\ref{tab:accuracy} we compare the accuracy of published models with our implementation on a Google dataset V1 \cite{DATA1} and V2 \cite{DATA2}.

We use the standard data set up from TensorFlow speech commands example code, proposed at \cite{DATA}. The NN is trained on twelve labels: ten words "yes", "no", "up", "down", "left", "right", "on", "off", "stop", and "go" with additional two words: “silence” and “unknown”. The "unknown" category contains remaining 20 keywords from the dataset. As in \cite{DATA, HELLO} we use an algorithm from \cite{CODE1} for splitting the data into training, validation and testing set with ratio 80:10:10. The length of the one training speech sample is 1 sec, and the sampling rate is 16kHz.

After applying the standard data set up (described above) on data sets V1 \cite{DATA1} we have 22246, 3093, 3081 samples for training validation and testing respectively. With data set V2 \cite{DATA2} we have 36923, 4445, 4890 samples for training validation and testing respectively.

The training data is augmented with:
\begin{itemize}[noitemsep]
\item time shift in range -100ms...100ms (as in \cite{CODE1}, \cite{HELLO});
\item signal resampling with resampling factor in range 0.85...1.15;
\item background noise (as in \cite{CODE1}, \cite{HELLO});
\item frequency/time masking, based on SpecAugment \cite{SPEC} (except time warping).
\end{itemize}

For side-by-side comparison purposes we use classification accuracy metric as in \cite{HELLO}. It is calculated by running the model on the testing data, and comparing the classification result against the expected label.

\subsection{Comparison with baseline}

We implemented popular KWS approaches DNN\cite{HELLO}, CNN\cite{HELLO}, LSTM\cite{HELLO}, GRU\cite{HELLO}, CRNN\cite{HELLO}, DSCNN\cite{HELLO}, TC-ResNet\cite{TEMPCONV}, described above, using our library for benchmarking streaming and non streaming models. We improved their accuracy on datasets V1 and V2, as shown on Table~\ref{tab:accuracy}, by applying SpecAugment \cite{SPEC}(except time warping) with hyper-parameters optimization of both neural net and speech feature extractor parameters. After model is trained on datasets V1/V2 we convert it to TFLite format (to be able to run it on a mobile phone), then run inference with TFLite and report its accuracy on Table~\ref{tab:accuracy} (columns V1, V2). The baseline accuracy with reference to paper is shown on Table~\ref{tab:accuracy} (columns V1*, V2*).

One of the best KWS models is Embed+head\cite{LIMSDATA}. It achieves the state of the art accuracy by using additional data sets from YouTube to train embedding layer. We introduced MHAtt-RNN model. It reduces classification error on datasets V2 by 10\% in comparison to Embed+head\cite{LIMSDATA}, as shown on Table~\ref{tab:accuracy}. The cost of this improvement is MHAtt-RNN model has two times more parameters than Embed+head\cite{LIMSDATA}. Another recently published promising approach is Matchbox\cite{MATCHBOX}, but authors use different training testing data set up, so we could not compare it side by side.

In addition we implemented and benchmarked SVDF model\cite{SVDF1} on both data sets V1 and V2 shown at Table~\ref{tab:accuracy} and demonstrated that it has good properties for streaming at Table~\ref{tab:accuracy_latency_size}.

\subsection{Streaming and non-streaming latency with accuracy}

In the real production environment we do not know neither the beginning nor ending of the speech command produced by the user. Also we need to provide real time responses for a good user experience. As a result, the speech command detector is running in streaming mode by classifying every 20 milliseconds (for example) of the input audio stream.

We trained all the models on datasets V2 and converted DNN, CNN no stride, CRNN, DSCNN no stride, SVDF to a streaming inference mode and benchmarked them on datasets V2. The models DSCNN with stride, CNN with stride and MHAtt-RNN are not streamable with our library. To emulate the streaming environment during accuracy evaluation we did not reset the RNN model states between testing sequences. We observed up to 2x accuracy reduction on such models (also baseline RNN models in Table~\ref{tab:accuracy} have the same issue). We addressed it by re-training RNN models  GRU, CRNN) with stateful argument=True\cite{STA}. The last state for each sample at index i in a batch will be used as initial state for the sample of index i in the following batch (the model will learn how to process cell states on its own). To distinguish such RNN models on Table~\ref{tab:accuracy_latency_size} we append (S). We observed accuracy reduction of statefully trained models GRU (S) CRNN (S), shown on Table~\ref{tab:accuracy_latency_size} in comparison to their non statefully trained versions GRU, CRNN shown Table~\ref{tab:accuracy} (column V2).

The latency and accuracy of non-streaming and streaming models with the number of parameters are presented in Table~\ref{tab:accuracy_latency_size}. We use a Pixel 4 mobile phone\cite{PIXEL4} and TFlite benchmarking tools \cite{BENCH} to measure the models latency. Non-streaming latency is the processing time of the whole 1 sec speech sequence by the non-streaming model representation. Streaming latency is the processing time of one audio frame by the streaming model (it receives 20ms of audio and returns classification result). Processing time includes both feature extraction and neural network classification (end to end).

In Table~\ref{tab:accuracy_latency_size} we observe that the most effective and accurate streaming models are SVDF, CRNN and GRU. Layers with striding/pooling are not streamable in our library now, but it can be implemented in the future. With support of striding/pooling in streaming mode, models such as TC-ResNet, Embed+head and Matchbox can be more preferable. The most accurate non-streaming model is MHAtt-RNN. It is based on bidirectional LSTM, so non streamable by default.

Table~\ref{tab:accuracy_latency_size} shows that the average latency ratio of non-streaming to streaming convolutional models (CNN, SVDF, DSCNN) is around 10x. The same ratio for RNN models (GRU(S), CRNN(S)) is around 20x. We explain this difference by the fact that non-streaming RNN models still have to be executed sequentially frame by frame over all frames belonging to 1 second of input audio, whereas non-streaming convolutions can be computed over the whole sequence in one batch. In an ideal case this ratio has to be around 50x (there are 50 frames in one second). As a result, there are opportunities for latency speed-up of streaming models: reduce memory allocations in the ring buffers and enable support of internal state in the inference engine. At the same time the latency of non-streaming models also can be optimized further. Detailed profiling of non-streaming models showed that the speech feature extraction latency is around 3.7ms. It can be reduced by using FFT and model quantization (after enabling both we observe almost 2x latency reduction). As expected non-streaming models with striding are several times faster than the same model with no striding: CNN and DSCNN with striding are two times faster than the same models without striding.

\begin{table}[t]
  \caption{Accuracy on data V2 with latency and model size}
  \label{tab:accuracy_latency_size}
  \centering
  \begin{tabular}{p{1.7cm} p{1.10cm} p{1.15cm} p{1.10cm} p{0.90cm}}
    \toprule
    \textbf{Models} &
    \textbf{accuracy, [\%]} &
    \textbf{non stream latency, [ms]} &
    \textbf{stream latency, [ms]} &
    \textbf{model size, [K]}  \\
    \midrule
      DNN        & 90.6  & 4  & 0.6  & 447 \\
      CNN+strd   & 95.6  & 6  & N/I  & 529 \\
      CNN        & 96.0  & 15 & 1.5  & 606 \\
      GRU (S)    & 96.3  & 11 & 0.5  & 593 \\
      CRNN (S)   & 96.5  & 9  & 0.5  & 467 \\
      SVDF       & 96.9  & 5  & 0.6  & 354 \\
      DSCNN      & 96.9  & 19 & 1.6  & 490 \\
      DSCNN+strd & 97.0  & 9  & N/I  & 485 \\
      TC-ResNet  & 97.4  & 5  & N/I  & 365 \\
      MHAtt-RNN  & 98.0  & 10 & N/A  & 743 \\
    \bottomrule
  \end{tabular}
  \vspace{-5mm} % TODO remove it
\end{table}

\section{Conclusion}

We built a library allowing end-to-end model conversion to streaming inference on mobile phones using Keras and TFLite. The converted model encapsulates speech feature extraction. It simplifies model deployment on mobile devices. We reduced classification error by 10\% relative, in comparison to the state of the art models on datasets V2. It was achieved by extending the Att-RNN\cite{ATT2} model with multi-head attention and applying SpecAugment \cite{SPEC}. For benchmarking purpose we implemented several popular models using our streaming library and measured streaming and non streaming latency on a Pixel 4 mobile phone and demonstrated different tradeoffs between accuracy and latency. All code with experimentation results are open-sourced and available at \cite{OPEN}.

\section{Acknowledgements}

The authors would like to thank Pete Warden, Karolis Misiunas, Yukun Zhu, Ben Vanik, Robert Suderman, Rohit Prabhavalkar, Kevin Kilgour and BDI team for valuable discussions and suggestions.

\Urlmuskip=0mu plus 1mu\relax
\bibliographystyle{IEEEtran}
\bibliography{kws_streaming_paper}

\end{document}